\documentclass[journal,twoside,web]{ieeecolor}
\usepackage{jsen}
\usepackage{cite}
\usepackage{amsmath,amssymb,amsfonts}
\usepackage[pdftex]{graphicx}
\usepackage{textcomp}
\usepackage{wrapfig}

\usepackage[hang,flushmargin]{footmisc}
\usepackage{float}
\usepackage{lipsum}
\usepackage{epstopdf}
\usepackage{bm}
\usepackage[noend]{algpseudocode}
\usepackage{algorithmicx,algorithm}

\usepackage{subfig}

\def\BibTeX{{\rm B\kern-.05em{\sc i\kern-.025em b}\kern-.08em
    T\kern-.1667em\lower.7ex\hbox{E}\kern-.125emX}}
\definecolor{abstractbg}{rgb}{0.89804,0.94510,0.83137}
\begin{document}
\title{Received Signal Strength Based Wireless Source Localization with Inaccurate Anchor Positions}
\author{Yang Liu, Guojun Han,\IEEEmembership{Senior Member, IEEE} and Yonghua Wang, \IEEEmembership{Member, IEEE}, Zheng Xue, Jing Chen, Chang Liu
\thanks{This work was supported in part by Natural Science Foundation of China (Nos. U2001203, 61871136, 62071131 ), Natural Science Foundation of Guangdong Province (2014A030310266).). }
\thanks{Yang Liu, Guojun Han, Zheng Xue and Chang Liu are with the School of Information Engineering, Guangdong University of Technology, GuangZhou, China, (email:liuyang@gdut.edu.cn; gjhan@gdut.edu.cn;  xuezheng@mail2.gdut.edu.cn;liuchang@gdut.edu.cn)}
\thanks{Yonghua Wang is with the School of Automation, Guangdong University of Technology, GuangZhou, China,(email:wangyonghua@gdut.edu.cn)}
\thanks{Jing Chen is with the School of Phyics\&Optoelectronic Engineering, Guangdong University of Technology, Guangzhou, China, (email:jchen125@gdut.edu.cn)}}

\IEEEtitleabstractindextext{%
\fcolorbox{abstractbg}{abstractbg}{%
\begin{minipage}{\textwidth}%
\begin{wrapfigure}[9]{r}{3in}%
\includegraphics[width=3in]{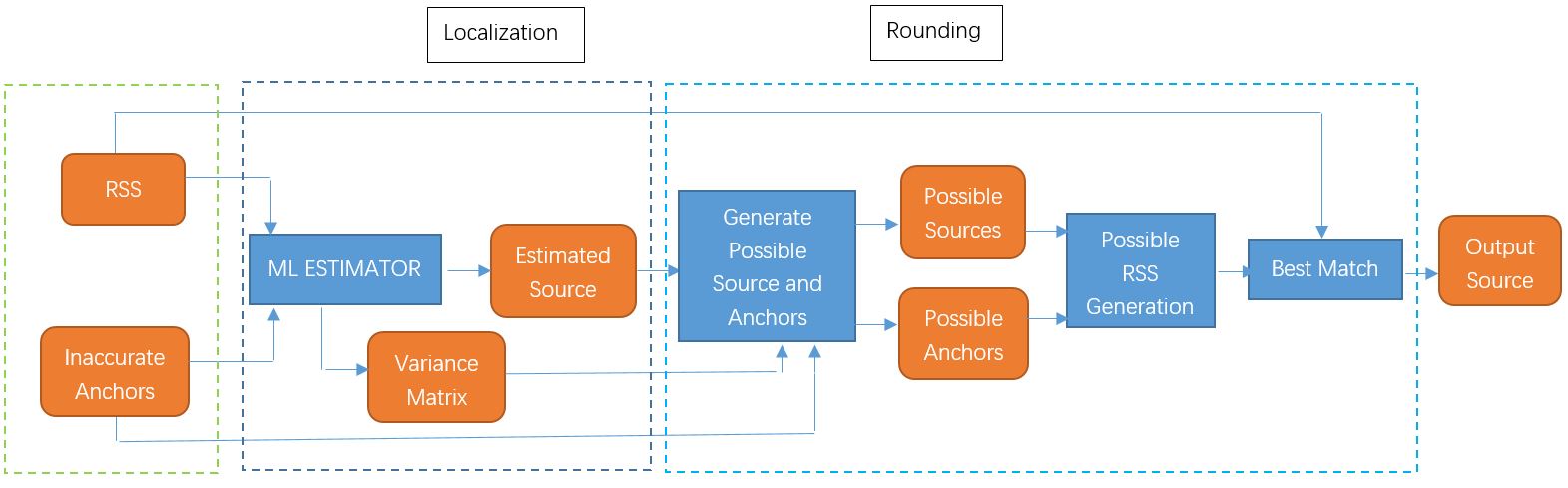}%
\end{wrapfigure}%
\begin{abstract}
Received signal strength (RSS)-based
wireless localization is easy to implement at low cost. In practice,
exact positions of anchors may not be available. This paper focuses
on determining the location of a source in the presence of inaccurate
positions of anchors based on RSS directly. We first use
Taylor expansion and a min-max approach to get a
maximum likelihood estimator of the coordinates of the source. Then
we propose a relaxed semi-definite programming model
to circumvent the non-convexity. We also propose a
rounding algorithm considering both inaccurate source locations
and inaccurate anchor locations.
Simulation results together
with analysis are presented to validate the proposed method.
\end{abstract}

\begin{IEEEkeywords}
Robust Source Localization; Received Signal Strength; Inaccurate Anchor Position; Semidefinite Programming
\end{IEEEkeywords}
\end{minipage}}}

\maketitle

\section{Introduction}
%
%
%

\IEEEPARstart{W}{ireless} source localization is a crucial aspect in many applications. Source localization aims at locating a target based on measurements related to pre-deployed distributed sensors with prior known locations, i.e., anchor nodes. These measurements include, for example, Time of Arrival (TOA), Time Difference of Arrival (TDOA), Angle of Arrival (AOA), and Received Signal Strength (RSS). The high feasibility, simplicity, and deployment practicability of RSS make it applicable for very simple devices with extremely constrained resources referred to as 'smart dust'\cite{Yu2021,Ouyang2021,Dinh2021,Jiang2021}. RSS based method requires neither clock synchronization nor an antenna array; so that it is more cost-effective in terms of both hardware and software.


A lot of research works assume that the known positions of anchors are accurate. However, in reality, this assumption often is too strong. Neglecting the inaccuracy of anchor position will undoubtedly deteriorate the localization performance severely. In addition, the accuracy of the localization algorithm will also be influenced by the measurement noise and the number of anchors. Hence, the localization algorithm should be 'robust' to overcome the anchor location uncertainty and measurement noise.

In this paper, we model the inaccuracy of the anchors as a bounded random vector. Based on this model we propose a robust method dealing with the inaccuracy of anchors. Firstly, we propose a maximum likelihood nonconvex model based on RSS. Then we transform and relax the original problem to an Semidefinite Programming Relaxation (SDR) problem, which is convex and easy to solve. Meanwhile, we propose a rounding algorithm, considering both inaccurate anchor positions and source positions. Additionally, we derive the number of times needed for the Monte Carlo method. Through comparison with several other techniques, simulation results show that the proposed method outperforms the others.

It should be mentioned here that localization problems with inaccurate anchor positions have been studied in previous research. However, these earlier works typically adopts a simpler assumption. For example, \cite{Angjelichinoski2014,Angjelichinoski2015} consider the estimated anchor positions to be normally distributed around the exact positions of anchors. How to perform source localization without having information of anchor positions remain to be explored.

%

The contributions of this paper are highlighted as follows:
\begin{enumerate}
    \item We propose a source location estimator using RSS measurement directly. This estimator is robust in application scenarios where locations of anchors are unavailable.
    \item We relax the proposed robust estimator to a Semidefinite Programming (SDP) problem so that it can be solved efficiently.
    \item We analyze different problem-dependent rounding strategies. Then we propose a rounding algorithm considering inaccurate anchors and non-feasible source location simultaneously.
\end{enumerate}

The paper is organized as follows. Section 2 reviews related works. Section 3 describes the basic idea and provides details of the proposed convex estimator for source localization using RSS with inaccuracy anchor locations. This section also analyzes different rounding strategies. Section 4 presents numerical and real experiment results. This section also gives analysis for the proposed localization method comparing with other methods elaborately. Finally, concluding remarks and future research directions are given in Section 5.

\emph{Notation}: Bold lower case letters are used to denote column vectors.

\section{Related Works}

Based on different types of physical measurements, source localization approaches mainly fall into several categories: RSS\cite{Yu2021,Ouyang2021,Dinh2021,Jiang2021,Dinh2020,Yang2020,Kumar2020}, TOA\cite{RN840}, TDOA\cite{RN843}, DOA\cite{RN850},etc. These physical measurements are typically used to estimate distances between the source and anchors to construct a distance matrix used for localization(e.g.\cite{liu2013local,chen2013wireless}). For example, Ref.\cite{Dinh2020} proposes a real-time indoor tracking and positioning system using Bluetooth Low Energy beacon and smartphone sensors. Based on the analysis of RSS, this paper presents a distance estimation method, and then estimates the initial positions through a trilateration technique. Ref.\cite{Yang2020} introduces a novel way of detecting drones by smartphones using RSS. Ref.\cite{Kumar2020} analyses the performance of target detection and localization methods in heterogeneous sensor networks using compartmental model, which is an attenuation model expressing the variation of RSS with propagation distance.

There are four commonly used types of estimators for source localization: Maximum Likelihood (ML), Least Square (LS), Semidefinite Programming (SDP), and Second-Order Cone Programming (SOCP). The ML and LS-based methods are highly nonconvex, so finding the global optimum is often with high computational complexity, especially for a large-scale problem. For the same reason, local optima (instead of the global optimum) might be obtained depending on the selected initial point. SDP and SOCP based methods deal with the non-convexity by relaxing the nonconvex constraints in the original problems so that they can be formed into convex ones. For these methods, the tightness of the relaxation shall be considered to guarantee accuracy. In addition, it is shown that though SOCP relaxation has a more straightforward structure and can be solved faster, its effectiveness is typically
weaker than the SDP method\cite{Tseng2007}.

Inaccurate anchor positions can deteriorate the localization performance. In recent years, there has been an increasing interest in determining source positions with inaccurate anchor positions\cite{RN872,RN874,RN870,RN869,Naddafzadeh-Shirazi2014,RN871,Hao2019,Angjelichinoski2014}\cite{Xiong2017}\cite{Angjelichinoski2015}\cite{Tomic2015}\cite{Denkovski2017}. Ref.\cite{RN872} proposes a min-max method for relative location estimation by minimizing worst-case estimation errors. The authors use the SDP technique to relax the original nonconvex problem into a convex one. Ref.\cite{RN874} focuses on differential received signal strength (DRSS)-based localization with model uncertainties such as unknown transmit power, Path Loss Element (PLE), and anchor location errors. This study presents a robust SDP-based estimator (RSDPE), which can cope with imperfect PLE and inaccurate anchor location information. Ref.\cite{RN870} performs analysis and develops a solution for locating a moving source using TDOA measurements in the presence of random errors in anchor locations. Ref.\cite{Naddafzadeh-Shirazi2014} proposes a mixed robust SDP-SOCP framework to benefit from the better accuracy of SDP and the lower complexity of SOCP. Ref.\cite{RN871} uses TDOA information considering the sensor node's (anchor) bounded location error effect. Ref.\cite{Hao2019} considers rigid body localization using the range measurements between the sensors on the body and the outside anchors that have position uncertainties, where a calibration emitter at an inaccurate location is employed to mitigate the anchor position errors. Ref.\cite{Xiong2017} studies performance limits of anchorless cooperative localization
in WSNs with strong sensor position uncertainty. This paper shows that the average localization performance of a network is largely determined by the number of agents and the signal metric employed rather than the network topology. Ref.\cite{Angjelichinoski2015} introduces an RSS-based framework for joint estimation of the positions of a wireless transmitter source and the corresponding measuring anchors. The framework exploits the imprecise anchor position information using non-Bayesian estimation and employs a novel Joint Maximum Likelihood (JML) algorithm for reliable anchor and agent position estimations. Ref.\cite{Tomic2015} proposes new approaches based on convex optimization to solve the received signal strength (RSS)-based noncooperative and cooperative localization problems in wireless sensor networks.Ref.\cite{Denkovski2017} jointly estimates unknown source and uncertain anchors positions and derives theoretical limits of the framework.

\section{Robust Localization Considering Inaccurate Anchor Positions}

\subsection{Problem Formulation}

The signal strength measurement is subject to a complicated radio propagation channel. In this paper, the log-normal shadowing model is used to characterize RSS from the source and received by the $i$-th anchor, which is denoted as $\Omega_i$\cite{Goldsmith2005}.
\begin{equation}\label{eq:RSS measurement model}
L_i=L_0+10\gamma\log_{10}\frac{||\mathbf{x}-\mathbf{z_i|}|}{d_0}+n_i
\end{equation}

where $\bm{\mathbf{x}}$ denotes the source position to be determined. $P_T$ denotes the transmit power of the source, $L_i=P_T-\Omega_i$ denotes the path loss, $n_i$ denotes a Gaussian random variable representing the noise of RSS measurement, $L_0$ denotes the path loss value at the reference distance $d_0$, and $\gamma$ denotes the path loss exponent.

Suppose that there are $M$ location-aware anchors and one location-unaware source. As mentioned above, in practice, the position information of anchors are typically inaccurate. Therefore, the true position $\mathbf{z_i}$  of anchor $i$ can be expressed as
\begin{equation}\label{eq:relationship between true and inaccurate}
\mathbf{z_i}=\hat{\mathbf{z_i}}+\mathbf{\Delta_i},\quad \text{for}\quad i=1,2,...M
\end{equation}
where $\hat{\mathbf{z_i}}$ denotes the inaccurate position information of anchor $i$, $\mathbf{\Delta_i}$ denotes the localization error, which is assumed to be bounded,
\begin{equation}\label{eq:error is bounded}
\|\mathbf{\Delta_i}\|\leq \zeta,\quad \text{for}\quad i=1,2,...M
\end{equation}
in \eqref{eq:error is bounded}, $\|\|$ denotes the Euclidean norm and $\zeta$ denotes the upper bound of the localization error.

From \eqref{eq:RSS measurement model}, the corresponding Maximum Likelihood estimator is
\begin{equation}\label{eq:ML estimator of target with inaccuracy of anchor}
\mathbf{\mathbf{x}}_P=\arg\min_\mathbf{\mathbf{x}}\sum_{i=1}^{M}\left(10\gamma\log_{10}\frac{||\mathbf{\mathbf{x}}-\mathbf{z_i}||}{d_0}-(L_i-L_0)\right)^2
\end{equation}
Let
\begin{equation}
\beta_i^2=d_0^2 10^{\frac{L_i-L_0}{5\gamma}}
\end{equation}
Then, the ML estimator \eqref{eq:ML estimator of target with inaccuracy of anchor} can be written as
\begin{equation}\label{eq:original problem for rounding}
\mathbf{\mathbf{x}}_P=\arg\min_\mathbf{\mathbf{x}}\sum_{i=1}^{M}\left(\log_{10}\frac{||\mathbf{\mathbf{x}}-\mathbf{z_i}||^2}{\beta_i^2}\right)^2 \end{equation}

Considering Eq.\eqref{eq:relationship between true and inaccurate}, an optimization approach is proposed:
\begin{equation}\label{eq:inaccurate anchor position model}
\begin{split}
&\mathbf{\mathbf{x}}_P=\arg\min_\mathbf{\mathbf{x}}\sum_{i=1}^{M}\left(\log_{10}\frac{||\mathbf{\mathbf{x}}-\mathbf{z_i}||^2}{\beta_i^2}\right)^2
\\ &s.t. \quad \mathbf{z_i}=\hat{\mathbf{z_i}}+\mathbf{\Delta_i}
\end{split}
\end{equation}

Using Eq.\eqref{eq:error is bounded}, we can rewrite \eqref{eq:inaccurate anchor position model} as a min-max optimization problem for the worst-case design:
\begin{equation}\label{eq:min-max model}
\begin{split}
  &\mathbf{\mathbf{x}}_P=\arg\min_\mathbf{\mathbf{x}}\max_{||\mathbf{\Delta_i}||\leq \zeta}\sum_{i=1}^{M}\left(\log_{10}\frac{||\mathbf{\mathbf{x}}-\mathbf{z_i}||^2}{\beta_i^2}\right)^2
  \\&s.t. \quad \mathbf{z_i}=\hat{\mathbf{z_i}}+\mathbf{\Delta_i}
\end{split}
\end{equation}

By applying Taylor expansion, the term $||\mathbf{\mathbf{x}}-\mathbf{z_i}||$ in Eq.\eqref{eq:min-max model} can be expanded as
\begin{equation}\label{eq:taylor expansion}
  ||\mathbf{\mathbf{x}}-\mathbf{z_i}||=||\mathbf{\mathbf{x}}-\hat{\mathbf{z_i}}||-\frac{\mathbf{\Delta_i}^T(\mathbf{\mathbf{x}}-\hat{\mathbf{z_i}})}{||\mathbf{\mathbf{x}}-\hat{\mathbf{z_i}}||}+o(||\mathbf{\Delta_i}||)
\end{equation}

Letting $\delta_i=\frac{\mathbf{\Delta_i}^T(\mathbf{\mathbf{x}}-\hat{\mathbf{z_i}})}{||\mathbf{\mathbf{x}}-\hat{\mathbf{z_i}}||}$, then
\begin{equation}\label{eq:error bound another form}
|\delta_i|\leq \zeta
\end{equation}

Using \eqref{eq:taylor expansion} and \eqref{eq:error bound another form}, Eq.\eqref{eq:min-max model} can be transformed into:
\begin{equation}\label{eq:min-max model another form}
  \mathbf{\mathbf{x}}_P=\arg \min_\mathbf{\mathbf{x}} \max_{|\delta_i|\leq \zeta}\sum_{i=1}^{M}\left(\log_{10}\frac{(||\mathbf{\mathbf{x}}-\hat{\mathbf{z_i}}||-\delta_i)^2}{\beta_i^2}\right)^2
\end{equation}

which can also be written as
\begin{equation}\label{eq:min-max model another form norm version}
  \mathbf{\mathbf{x}}_P=\arg \min_\mathbf{\mathbf{x}} \max_{|\delta_i|\leq \zeta}\biggl|\biggl|\log_{10}\frac{(||\mathbf{\mathbf{x}}-\hat{\mathbf{z_i}}||-\delta_i)^2}{\beta_i^2}\biggl|\biggl|^2
\end{equation}

\subsection{The Relaxation Procedure}
In \eqref{eq:min-max model another form norm version}, $\{\mathbf{\mathbf{x}}:||\mathbf{\mathbf{x}}-\hat{\mathbf{z_i}}||=\delta_i\}$ is not in the domain of the objective function. Also, the source can not overlap with the anchors, so each anchor is a singular point. It is difficult to obtain and confirm the global minimum solution of \eqref{eq:min-max model another form norm version} because \eqref{eq:min-max model another form norm version} is obviously not convex.

To obtain a convex formulation, problem \eqref{eq:min-max model another form norm version} is transformed and relaxed, as shown in the following steps.

To facilitate the design of a convex estimator, we replace $||\cdot||^2$ in \eqref{eq:min-max model another form norm version} by $l_{\infty}$-norm , then we have

\begin{equation}\label{eq:chebychev norm approximation}
\begin{split}
  &\mathbf{\mathbf{x_P}}=\arg \min_\mathbf{\mathbf{x}}\max_i \biggl|\log_{10}\frac{(||\mathbf{\mathbf{x}}-\hat{\mathbf{z_i}}||-\delta_i)^2}{\beta_i^2}\biggl|
  \\&s.t. \quad |\delta_i| \leq \zeta
\end{split}
\end{equation}

By noting that

\begin{equation}\label{eq:remove abs}
\begin{split}
&\biggl|\log_{10}\frac{(||\mathbf{\mathbf{x}}-\hat{\mathbf{z_i}}||-\delta_i)^2}{\beta_i^2}\biggl|\\
&=\max\bigl(\log_{10}\frac{(||\mathbf{\mathbf{x}}-\hat{\mathbf{z_i}}||-\delta_i)^2}{\beta_i^2},
\log_{10}\frac{\beta_i^2}{(||\mathbf{\mathbf{x}}-\hat{\mathbf{z_i}}||-\delta_i)^2}\bigl)
\end{split}
\end{equation}

Since $\log_{10}(\mathbf{\mathbf{x}})$ is a strictly monotonically increasing function in its domain $(0,+\infty)$, we can rewrite Eq.\eqref{eq:chebychev norm approximation} as

\begin{equation}\label{eq:min-max remove abs and log version}
\begin{split}
\mathbf{\mathbf{x_P}}&=\arg\min_\mathbf{\mathbf{x}}\max_i\bigl(\log_{10}\frac{(||\mathbf{\mathbf{x}}-\hat{\mathbf{z_i}}||+\zeta)^2}{\beta_i^2},
\\&\log_{10}\frac{\beta_i^2}{(||\mathbf{\mathbf{x}}-\hat{\mathbf{z_i}}||-\zeta)^2}\bigl)
\\&=\arg\min_\mathbf{\mathbf{x}}\max_i\log_{10}\bigl(\frac{(||\mathbf{\mathbf{x}}-\hat{\mathbf{z_i}}||+\zeta)^2}{\beta_i^2},\frac{\beta_i^2}{(||\mathbf{\mathbf{x}}-\hat{\mathbf{z_i}}||-\zeta)^2}\bigl)
\\&=\arg\min_\mathbf{\mathbf{x}}\max_i\bigl(\frac{(||\mathbf{\mathbf{x}}-\hat{\mathbf{z_i}}||+\zeta)^2}{\beta_i^2},\frac{\beta_i^2}{(||\mathbf{\mathbf{x}}-\hat{\mathbf{z_i}}||-\zeta)^2}\bigl)
\end{split}
\end{equation}

Here we assume that $||\mathbf{\mathbf{x}}-\hat{\mathbf{z_i}}||-\zeta>0$.  This is reasonable since the inaccuracy of anchor's position is relatively small compared with the distances between source and anchors. Considering the 'worst-case' situation, we can just substitute $\delta_i$ by $\zeta$. Because Eq. \eqref{eq:min-max remove abs and log version} is still not convex, we introduce an auxiliary variable $k \in \mathbb{R}^+$. Then Eq.\eqref{eq:min-max remove abs and log version} can be transformed into
\begin{equation}\label{eq:add ausxiliary varialbe version}
\begin{split}
&\mathbf{\mathbf{x_P}}=\arg\min_{\mathbf{\mathbf{x}},k}k
\\&s.t. \quad \frac{(||\mathbf{\mathbf{x}}-\hat{\mathbf{z_i}}||+\zeta)^2}{\beta_i^2} \leq k \quad i=1,\ldots,M
\\&s.t. \quad \frac{\beta_i^2}{(||\mathbf{\mathbf{x}}-\hat{\mathbf{z_i}}||-\zeta)^2} \leq k \quad i=1,\ldots,M
\end{split}
\end{equation}

By noting that
\begin{equation}\label{eq:expand quadratic form}
\begin{split}
(||\mathbf{\mathbf{x}}-\hat{\mathbf{z_i}}||+\zeta)^2&=||\mathbf{\mathbf{x}}-\hat{\mathbf{z_i}}||^2+2\zeta||\mathbf{\mathbf{x}}-\hat{\mathbf{z_i}}||+\zeta^2
\\(||\mathbf{\mathbf{x}}-\hat{\mathbf{z_i}}||-\zeta)^2&=||\mathbf{\mathbf{x}}-\hat{\mathbf{z_i}}||^2-2\zeta||\mathbf{\mathbf{x}}-\hat{\mathbf{z_i}}||+\zeta^2
\end{split}
\end{equation}

and
\begin{equation}\label{eq:expand norm}
||\mathbf{\mathbf{x}}-\hat{\mathbf{z_i}}||^2=\mathbf{\mathbf{x}}^Tx-2\mathbf{\mathbf{x}}^T\hat{\mathbf{z_i}}+\hat{\mathbf{z_i}}^T\hat{\mathbf{z_i}}
\end{equation}
we can rewrite Eq.\eqref{eq:expand norm} as

\begin{equation}\label{eq:expand norm trace version}
||\mathbf{\mathbf{x}}-\hat{\mathbf{z_i}}||^2= \mathrm{tr}(X)-2\mathbf{\mathbf{x}}^T\hat{\mathbf{z_i}}+\hat{\mathbf{z_i}}^T\hat{\mathbf{z_i}}
\end{equation}
Where $\mathrm{tr}(X)$ denotes the trace of the auxiliary variable $X=xx^T \in \mathbb S^2$.

We then introduce an auxiliary variable $\mathbf{l}\in \mathbb{R}^{M\times 1}$, where $l_i=||\mathbf{\mathbf{x}}-\hat{\mathbf{z_i}}||\quad i=1,\ldots,M$. Letting $L=ll^T$, then

\begin{equation}
\label{eq:Lmatrix}
  L(i,i)=\mathrm{tr}(X)-2\mathbf{\mathbf{x}}^T\hat{\mathbf{z_i}}+\hat{\mathbf{z_i}}^T\hat{\mathbf{z_i}}
\end{equation}
It can be easily proved that,
\begin{equation}\label{eq:L element contrstraint ij}
  L(i,j)\geq 0
\end{equation}

Incorporating \eqref{eq:expand quadratic form} \eqref{eq:expand norm trace version} \eqref{eq:Lmatrix} \eqref{eq:L element contrstraint ij} into \eqref{eq:add ausxiliary varialbe version}, a formulation modified from \eqref{eq:add ausxiliary varialbe version} is obtained as
\begin{equation}\label{eq:revision of min-max model with rank one constraint}
\begin{split}
&\mathbf{\mathbf{x}}_P=\arg\min_{\mathbf{\mathbf{x}},k,l,X,L}k
\\&s.t.
\\&\mathrm{tr}(X)+2\mathbf{\mathbf{x}}^T\hat{\mathbf{z_i}}+\hat{\mathbf{z_i}}^T\hat{\mathbf{z_i}}+2\zeta l_i+\zeta^2 \leq k\beta_i^2
\\&\mathrm{tr}(X)+2\mathbf{\mathbf{x}}^T\hat{\mathbf{z_i}}+\hat{\mathbf{z_i}}^T\hat{\mathbf{z_i}}-2\zeta l_i+\zeta^2 \geq k^{-1}\beta_i^2
\\& L(i,i)=\mathrm{tr}(X)-2\mathbf{\mathbf{x}}^T\hat{\mathbf{z_i}}+\hat{\mathbf{z_i}}^T\hat{\mathbf{z_i}}
\\& L(i,j)\geq 0
\\&X=\mathbf{x}\mathbf{x}^T
\\&L=\mathbf{l}\mathbf{l}^T
\\&k\geq 0
\\&i,j=1,\ldots,M
\end{split}
\end{equation}
In Eq.\eqref{eq:revision of min-max model with rank one constraint},      $\mathrm{tr}(X)+2\mathbf{\mathbf{x}}^T\hat{\mathbf{z_i}}+\hat{\mathbf{z_i}}^T\hat{\mathbf{z_i}}+2\zeta l_i+\zeta^2 \leq k\beta_i^2 $ are affine constraints. $\mathrm{tr}(X)+2\mathbf{\mathbf{x}}^T\hat{\mathbf{z_i}}+\hat{\mathbf{z_i}}^T\hat{\mathbf{z_i}}-2\zeta l_i+\zeta^2 \geq k^{-1}\beta_i^2 $ are convex constraints since $\mathrm{tr}(X)$ is linear in $X$, $\mathbf{\mathbf{x}}^T\hat{\mathbf{z_i}}$ is linear in $\mathbf{\mathbf{x}}$ and $k^{-1}$ is convex in $k$. $X=\mathbf{x}\mathbf{x}^T$ and $L=\mathbf{l}\mathbf{l}^T$ mean that $X$ and $L$ are rank one symmetric positive semidefinite (PSD) matrix, which indicates that $X\succeq 0,\quad \mathrm{rank}(X)=1$ and $L\succeq 0,\mathrm{rank}(L)=1$. It can be noticed that the fundamental difficulty in solving Eq.\eqref{eq:revision of min-max model with rank one constraint} is the set of the rank one constraint matriices, which is nonconvex. The objective function and all other constraints are convex in $\mathbf{\mathbf{x}},k,\mathbf{l},X,L$. Based on the above analysis, we relax the constraints $X=\mathbf{x}\mathbf{x}^T$ and $L=\mathbf{l}\mathbf{l}^T$ to $X\succeq \mathbf{x}\mathbf{x}^T$ and $L\succeq \mathbf{l}\mathbf{l}^T$, respectively. Using Schur complement, we have
\begin{equation}\label{eq:relax rank one constraint}
\begin{split}
  X\succeq \mathbf{x}\mathbf{x}^T \Rightarrow \begin{pmatrix}X&\mathbf{\mathbf{x}} \\ \mathbf{\mathbf{x}}^T& 1\end{pmatrix}\succeq 0
\\L\succeq \mathbf{l}\mathbf{l}^T \Rightarrow \begin{pmatrix}L&\mathbf{l} \\ \mathbf{l}^T& 1\end{pmatrix}\succeq 0
\end{split}
\end{equation}
Using Schur complement, we can also express $\mathrm{tr}(X)+2\mathbf{\mathbf{x}}^T\hat{\mathbf{z_i}}+\hat{\mathbf{z_i}}^T\hat{\mathbf{z_i}}-2\zeta l_i+\zeta^2 \geq k^{-1}\beta_i^2$ in a Linear Matrix Inequality (LMI) form as follow:
\begin{equation}\label{LMI expression of t constraint}
\begin{split}
&\begin{pmatrix} \mathrm{tr}(X)+2\mathbf{\mathbf{x}}^T\hat{\mathbf{z_i}}+\hat{\mathbf{z_i}}^T\hat{\mathbf{z_i}}-2\zeta l_i+\zeta^2 & \beta_i\\ \beta_i &k \end{pmatrix} \succeq 0\\
      &       i=1,\ldots,M
\end{split}
\end{equation}
Incorporating Eq.\eqref{eq:relax rank one constraint}, Eq.\eqref{LMI expression of t constraint} into Eq.\eqref{eq:revision of min-max model with rank one constraint}, the received signal strength based robust source localization problem can be described as follows:
\begin{equation}\label{eq:convex form}
\begin{split}
&\mathbf{\mathbf{x}}_P=\arg\min_{\mathbf{\mathbf{x}},k,l,X,L}k
\\&s.t.
\\&\mathrm{tr}(X)+2\mathbf{\mathbf{x}}^T\hat{\mathbf{z_i}}+\hat{\mathbf{z_i}}^T\hat{\mathbf{z_i}}+2\zeta l_i+\zeta^2 \leq k\beta_i^2
\\&\begin{pmatrix} \mathrm{tr}(X)+2\mathbf{\mathbf{x}}^T\hat{\mathbf{z_i}}+\hat{\mathbf{z_i}}^T\hat{\mathbf{z_i}}-2\zeta l_i+\zeta^2 & \beta_i\\ \beta_i &k \end{pmatrix} \succeq 0
\\ &L(i,i)=\mathrm{tr}(X)-2\mathbf{\mathbf{x}}^T\hat{\mathbf{z_i}}+\hat{\mathbf{z_i}}^T\hat{\mathbf{z_i}}
\\ &L(i,j)\geq 0
\\&\begin{pmatrix}X&\mathbf{\mathbf{x}} \\ \mathbf{\mathbf{x}}^T& 1\end{pmatrix}\succeq 0
\\&\begin{pmatrix}L&\mathbf{l} \\ \mathbf{l}^T& 1\end{pmatrix}\succeq 0
\\&k\geq 0
\\&i,j=1,\ldots,M
\end{split}
\end{equation}
In Eq.\eqref{eq:convex form}, $\mathbf{\mathbf{x}}\in \mathbb{R}^2$ represents the position of interest. $\mathbf{\mathbf{x}}_P$ represents the corresponding estimator of $\mathbf{\mathbf{x}}$. The remaining variables are $k\in\mathbb{R}$, $\mathbf{l}\in \mathbb{R}^M$, $X\in \mathbb{R}^{M\times M}$, $L \in \mathbb{R}^{M\times M}$. All constraints in Eq.\eqref{eq:convex form} are expressed as LMIs. Eq.\eqref{eq:convex form} is an instance of semidefinite programming (SDP) and also a relaxation of Eq.\eqref{eq:add ausxiliary varialbe version}.
\subsection{Rounding the Solution}

A fundamental issue that one must address when using SDR is how to round the solution. In this paper the word 'round' has two implications. Firstly, we should convert a globally optimal solution $\mathbf{\mathbf{x}}^*$ of problem Eq.\eqref{eq:convex form} into a feasible solution $\tilde{\mathbf{x}}$ of problem Eq.\eqref{eq:revision of min-max model with rank one constraint}. Secondly, from the solution of Eq.\eqref{eq:convex form}, the rounding procedure can generate candidates 'nearby' and select the 'best match' of the original problem's constraints.

An intuitively appealing idea is to apply the rank-one approximation on $X^*$, $L^*$, $\mathbf{\mathbf{x}}^*$ and $\mathbf{l}^*$ of Eq.\eqref{eq:convex form}. This method uses the largest eigenvalue $\lambda_{xM}$ and the corresponding eigenvector $q_{xM}$ to approximate $X^*$, uses $\lambda_{lM}$ and the corresponding eigenvector $q_{lM}$ to approximate $L^*$. Then if $\sqrt{\lambda_{xM}}q_{xM}$  and $\sqrt{\lambda_{lM}}q_{lM}$ are feasible to Eq.\eqref{eq:revision of min-max model with rank one constraint}, we get the solution $\tilde{\mathbf{x}}$. Otherwise $\sqrt{\lambda_M}q_M$ and $\sqrt{\lambda_{lM}}q_{lM}$ need to be mapped to nearby feasible solutions in a problem dependent way.

It is also convenient and straightforward to use the solution of Eq.\eqref{eq:convex form} as an initial point for the original problem Eq.\eqref{eq:original problem for rounding} and run a local optimization method. However, Eq.\eqref{eq:original problem for rounding} is non-differentiable and not completely smooth. In contrast, Eq.\eqref{eq:revision of min-max model with rank one constraint} is smooth and asymptotic optimal. However, in Eq.\eqref{eq:revision of min-max model with rank one constraint}, the constraints are complicated so that it is hard to compute the Newton step.

Randomization is another way to extract an approximate solution from an SDR solution. The key of this method is to use $X^*-\mathbf{\mathbf{x}}^*{\mathbf{\mathbf{x}}^*}^T$ of Eq.\eqref{eq:convex form} as a covariance matrix. We generate random vectors $\xi_\mathbf{\mathbf{x}}\sim N(0,X^*) $, and then use the random vector to construct an approximate solution in a problem-dependent way. The rounding algorithm using randomization is described in Alg.\ref{alg:refine solution}.
\begin{algorithm}[htbp]
\caption{Refine By Randomization}\label{alg:refine solution}
\label{alg:randomization}
\begin{algorithmic}[1]
\Require $\mathbf{\mathbf{x}}^*$,$X^*$,$tt$,$M$
\Ensure $\mathbf{\mathbf{x}}$
\For{$i=1:\text{tt}$}
\State $\mathbf{\mathbf{x}}^o(i) \sim N(\mathbf{\mathbf{x}}^*,X^*-\mathbf{\mathbf{x}}^*{\mathbf{\mathbf{x}}^*}^T)$
\State $k_o(i) \gets \text{Computek}(\mathbf{\mathbf{x}}^o(i),M)$
\EndFor
\State $i \gets \arg \min_i k_o(i)$
\State $\mathbf{\mathbf{x}} \gets \mathbf{\mathbf{x}}^o(i)$
\end{algorithmic}
\end{algorithm}

\begin{algorithm}[htbp]
\caption{Create Feasible Solution}\label{alg:create feasible solution}
\begin{algorithmic}[1]
\Procedure{Computek}{$\mathbf{\xi}$, M}
\State $X^o \gets \mathbf{\xi}_x\mathbf{\xi}_x^T$
\For{i=1:M}
${l}^o_i \gets ||\mathbf{x}^o-\hat{\mathbf{z_i}}||$
\EndFor
\State $L^o \gets [{l}^o_i]_{i=1:M}$
\State Compute $k^o$ using $X^o$, $L^o$, $\xi$, ${l}^o$ in Eq.\eqref{eq:revision of min-max model with rank one constraint}
\EndProcedure
\end{algorithmic}
\end{algorithm}

In Alg.\ref{alg:refine solution}, to generate $\xi_\mathbf{\mathbf{x}}(t)$, we simply generate a random vector $\mathbf{u}$ whose components follow ii.d $N(0,1)$, then let $\mathbf{\xi}_x(t)=V^T\mathbf{u}+\mathbf{\mathbf{x}}^*$, where $V$ is the factorization matrix $X^*-\mathbf{\mathbf{x}}^*{\mathbf{\mathbf{x}}^*}^T=V^TV$. Since $X^*-\mathbf{\mathbf{x}}^*{\mathbf{\mathbf{x}}^*}^T \succeq 0$ , $V$ always exists.  In, Alg.\ref{alg:refine solution}, Alg.\ref{alg:create feasible solution} is called to create a feasible solution of Eq.\eqref{eq:revision of min-max model with rank one constraint}.


Another approach is through grid search. Grid search has a significantly reduced search space which allows it to find the feasible solution quickly and accurately. For a Gaussian random vector $\mathbf{\xi}_x(t)$ generated in Alg.\ref{alg:refine solution}, it exists close to the mean with a high probability. Then the randomization method wastes a certain number of trials in the 'sparse' area. It is also reasonable to adapt a variable step size to balance searching performance and complexity. In this paper, an adaptive variable step size based grid searching rounding algorithm is proposed in Alg.\ref{alg:grid searching}. Note that in Alg.\ref{alg:grid searching} a mechanism is designed to reduce the searching space when the candidate point $\mathbf{\mathbf{x}}^o$ is far from the mean $\mathbf{\mathbf{x}}^*$. In addition, Alg.\ref{alg:grid searching} restricts the searching scope to $3*\max([X^*]_{1,1}, [X^*]_{2,2})$. Considering a one-dimensional situation, given a Gaussian random variable $\mathbf{\mathbf{x}}$, $P (||\mathbf{\mathbf{x}}-E(\mathbf{\mathbf{x}})|| \leq 3 \sigma_x) \geq 0.999$. This demonstrates that the searching scope is sufficiently large to find the best candidate point.
\begin{algorithm}[htbp]
\caption{Refine By Variable Step Grid Searching}\label{alg:grid searching}
\begin{algorithmic}[1]
\Require $\mathbf{\mathbf{x}}^*$,$X^*$
\Ensure $\mathbf{\mathbf{x}}$
\State $\mathbf{\mathbf{x}}^o \gets \mathbf{\mathbf{x}}^*$
\State $\sigma_d \gets \max([X^*]_{1,1}, [X^*]_{2,2})$
\State $d_s \gets 0.0001$
\State $\Delta_{ds} \gets 0.001*\sigma_d$
\While{$d_s=|| \mathbf{\mathbf{x}}^o-\mathbf{\mathbf{x}}^* ||) \leq 3\sigma_d $}\label{algline:probability range}
    \State $\iota=\text{fix}(\frac{1}{d_s})$
    \State $[\Delta_\mathbf{\mathbf{x}}]_{1...\iota} \gets [d_s*\cos(2\pi*\frac{1...\iota}{\iota})]$
    \State $[\Delta_y]_{1...\iota} \gets [d_s*\sin(2\pi*\frac{1...\iota}{\iota})]$
    \State $[\xi]_{1...\iota} \gets \mathbf{\mathbf{x}}^o+[\Delta_\mathbf{\mathbf{x}},\Delta_y]^T$
    \State $k_o \gets \text{Computek}(\xi,M)$
    \State $d_s \gets d_s+\Delta_{ds}$
\EndWhile
\State $t \gets \arg \min_t k_o(t)$
\State $\mathbf{\mathbf{x}} \gets \xi_x(t)$
\end{algorithmic}
\end{algorithm}

\begin{figure*}[htbp]%

\centering

\subfloat[][]{\includegraphics[width=0.45\linewidth]{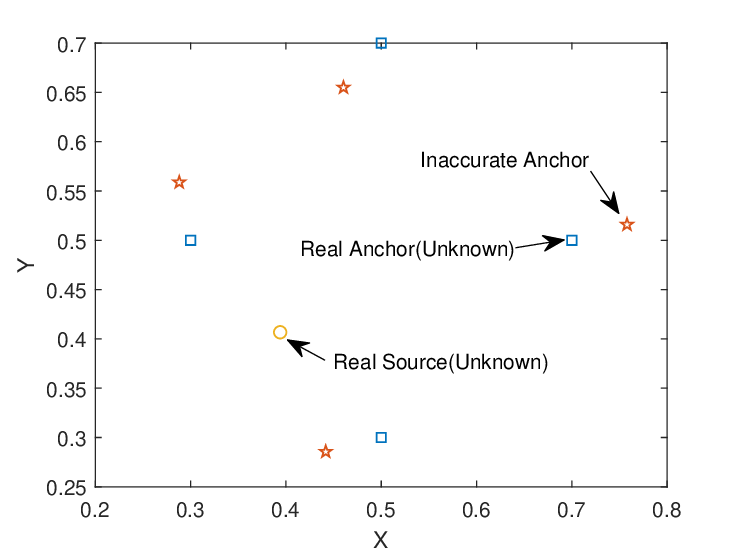}\label{sub:illus1}}%
\subfloat[][]{\includegraphics[width=0.45\linewidth]{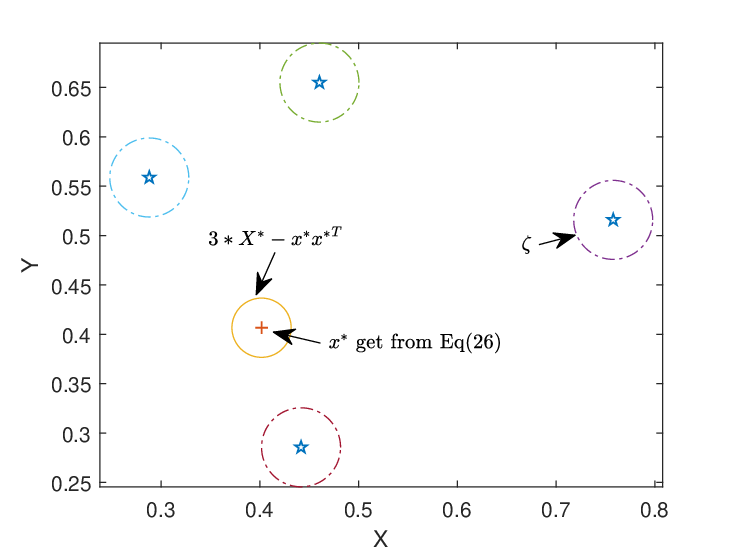}\label{sub:illus2}}\\
\subfloat[][]{\includegraphics[width=0.45\linewidth]{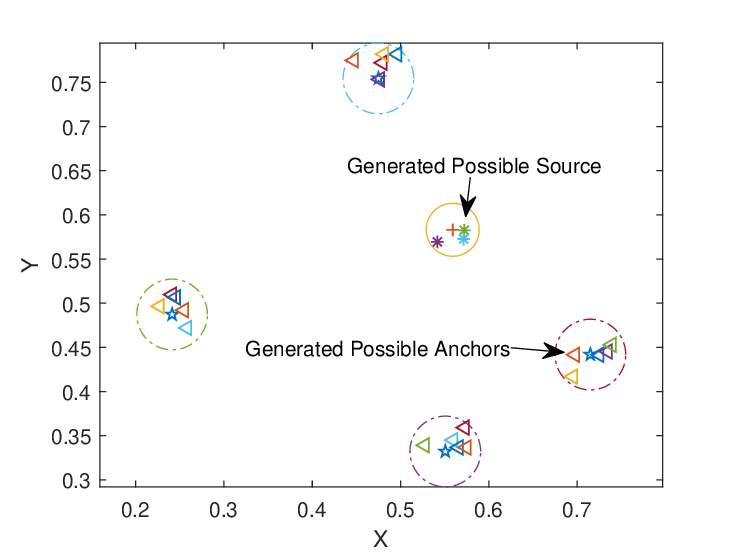}\label{sub:illus3}}
\subfloat[][]{\includegraphics[width=0.45\linewidth]{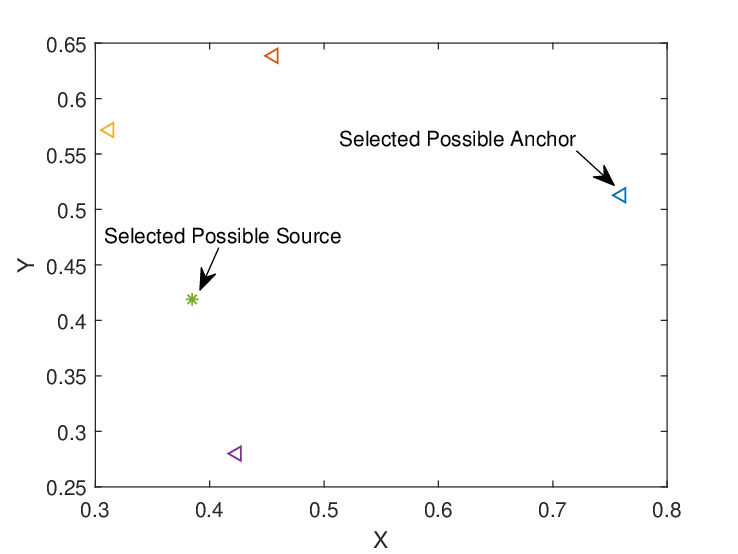}\label{sub:illus4}}\\
\subfloat[][]{\includegraphics[width=0.7\linewidth]{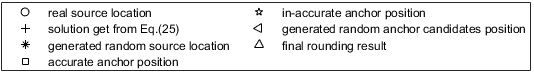}}%
\caption{Illustration of rounding algorithm Alg.\ref{alg:rr}. The circle denotes the real source location. Plus sign denotes the solution get from Eq.\eqref{eq:convex form}. The asterisk denotes the generated random source location. Square denotes the accurate anchor position. Pentagram denotes the in-accurate anchor position. Left-pointing triangle denotes the generated random anchor candidates position. Upward-pointing triangle denotes the final rounding result. Note that in this figure, the Upward-pointing triangle does not certainly represent the asterisk closest to the circle. This phenomenon is reasonable because the r-r method (actually, all rounding methods) only works on average.}%
\label{fig:illus r-r}%
\end{figure*}

Such rounding approaches, however, have failed to address anchor inaccuracy. To overcome this drawback, we propose a new rounding algorithm Alg.\ref{alg:rr}. This proposed algorithm takes two types of deviations into consideration simultaneously. One deviation is between the solution $\mathbf{x}^*$ of Eq.\eqref{eq:convex form} and the real source location. The other is between the unknown real anchor position $\mathbf{z}$ and the inaccurate anchor position $\mathbf{\hat{z}}$. It should be noticed that the power loss vector $L$ is accurate and reliable. In numerical simulation, $\mathbf{L}$ is generated using accurate anchor position $\mathbf{z}$ and the accurate source position $\mathbf{x}$ to simulate the reality. Alg.\ref{alg:rr} uses $\mathbf{L}$ to calculate the 'best match' of the source location and anchor position simultaneously. Fig.\ref{fig:illus r-r} illustrates the basic idea of Alg.\ref{alg:rr}.

Fig.\ref{sub:illus1} is the original hypothesis. There are four anchors with inaccurate position information and only the upper bound for the localization error $\zeta$ is available. Fig.\ref{sub:illus2} illustrates the situation after calculating \eqref{eq:convex form}. Here the approximate position $\mathbf{\mathbf{x}}^*$ and $\zeta$ are available, and the solution $X^*$ together with $\mathbf{\mathbf{x}}^*$ can provide the information of the real source location distribution as analyzed before. In Fig.\ref{sub:illus3}, Alg.\ref{alg:rr} generates a number of possible source locations denoted by asterisk distribute inside the circle whose radius can be calculated as $3(X^*-\mathbf{\mathbf{x}}^*{\mathbf{\mathbf{x}}^*}^T)$. In this subfigure, near every given anchor position $\hat{\mathbf{z_i}}$ (inaccurate), we generate several possible candidate anchor positions (denoted by the left-pointing triangle) within the range of $\zeta$. This step is described in Alg.\ref{alg:randomization}. In this subfigure, we calculate the RSS $L_{ij}^k \in \mathbb{R}$ using Eq.\eqref{eq:RSS measurement model}. Here $L_{ij}^k$ denotes the RSS between $\mathbf{\tilde{z}}_{ij}$ and $\mathbf{s}_k$.  $\mathbf{\tilde{z}}_{ij}, i=1 \ldots M, j=1 \ldots N$ denotes the $j$th possible anchor position near $\hat{z}_i$. $\mathbf{s}_k,k=1\ldots N$ denotes the $k$th possible source location. We then can formulate a RSS vector $\mathbf{L}_c^k \in \mathbb{R}^{M \times 1}$, where $\mathbf{L}_c^k(i)$ is the RSS received at the $i$th possible anchor position from the $k$th possible source position. Here $c$ denotes a combination of $j$, for example, $\{2,1,1,3,4\}$ and the like. Apparently $c$ is an element of a set $P$, where $P$ is the permutation of 1 to $N$. Here $N$ denotes the number of possible anchor positions we generated. Fig.\ref{sub:illus4} shows the final selected combination of the possible source and anchor locations, which has the lowest sum of RSS square error among all the possible combinations.

\begin{algorithm}[htbp]
\caption{Refine considering Inaccurate Anchors}\label{alg:rr}
\begin{algorithmic}[1]
\Require $\mathbf{\mathbf{x}}^*$,$X^*$, $tt$, $\zeta$, $\mathbf{\hat{z}}$, $M$, $N$,$\mathbf{L}$ \Comment\footnotemark[1]
\Ensure $\mathbf{\mathbf{x}}$
\For{$i=1:\text{tt}$}
\State $\mathbf{\mathbf{x}}^o(i) \sim N(\mathbf{\mathbf{x}}^*,X^*-\mathbf{\mathbf{x}}^*{\mathbf{\mathbf{x}}^*}^T)$
\EndFor
\For {$i=1:M$}
\For{$j=1:N$}
\State $\mathbf{\tilde{z}}(i,j) \sim U(G), G=\{\mathbf{\mathbf{x}}|||\mathbf{\mathbf{x}}-\mathbf{\hat{z}}_i|| \leq \zeta\}$ \Comment\footnotemark[3]
\EndFor
\EndFor
\State $p \gets \binom{M}{N}$ \Comment\footnotemark[4]
\For {$i=1:\text{tt}$}
\For {$j=1:\text{size}(p,2)$}
\State $r(i,j) \gets  \text{sre}(\mathbf{\mathbf{x}}^o(i),\mathbf{\tilde{z}}(p(j,:),L)$ \Comment\footnotemark[5]
\EndFor
\EndFor
\State $(i,j) \gets \min_{ij} r(i,j)$
\State $\mathbf{\mathbf{x}} \gets \mathbf{\mathbf{x}}^o(i)$
\end{algorithmic}
\end{algorithm}

It should be noted that 'r-r' rounding method can only be applied to SDP relaxation and cannot be applied to other types of relaxation (e.g., SOCP) for Eq.\eqref{eq:revision of min-max model with rank one constraint}. This is because without information about the variance matrix to guide the random solution generating procedure, random generation of feasible points can lead the rounding process impossible to compute.

\footnotetext[1]{$\mathbf{\mathbf{x}}^*$ and $X^*$ denote the solution of Eq.\eqref{eq:convex form}. $\zeta,\hat{z}, M, L$ are known parameters. $\mathbf{\hat{z}}\in R^{2\times M}$ is the matrix concatenating all inaccurate positions of the anchors $\mathbf{\hat{z}}_i, i=1 \ldots M$.  $tt$ denotes the number of generated possible source positions. $N$ denotes the number of generated possible anchor positions.}
\footnotetext[2]{$\mathbf{\mathbf{x}}$ denotes the true solution after rounding.}
\footnotetext[3]{$U(G)$ denotes the uniform distribution on area $G$.}
\footnotetext[4]{$\binom{M}{N}$ denotes the $M$-permutation of $N$. So that $p\in k \times M, k=N(N-1)(N-2)\ldots(N-M+1)$ is the set of all permutations. That is, every row of $p$ indicates a selected combination of the possible anchor positions.}
\footnotetext[5]{The function $\text{sre}$ computes the sum square error of RSS. That is, using Eq.\eqref{eq:RSS measurement model}, $r(i,j)=\sum_{k=1}^M((L_0+10\gamma\log_{10}\frac{||\mathbf{\mathbf{x}}^o(i)-\mathbf{\tilde{z}}(p(j,k))||}{d_0}+n_k)-L(j))^2$.}

\section{Simulation Results and Analysis}
Numerical simulations are performed in this section.  We conduct  Monte Carlo simulation to average out the effects of the geometric layout and random noise and error. The minimum number of trials needed is derived in the supplementary material. For each Monte Carlo trial, every anchor location is given with a random but bounded deviation. This paper evaluates the Root Mean Square  Error (RMSE) as the performance metric.

We parallel each random walk using the MATLAB command "parfor-loop".  Additional code optimization and C implementation are expected to further reduce the CPU time needed.


We use CVX\cite{Grant2014}  for specifying the convex problem, with SDPT3 \cite{Tuetuencue2003} as the solver. Note that currently CVX can not be used in the parallel loop. We circumvent this problem by defining a function that contains the CVX code and then call it in the 'parfor' loop.

We use modified ML estimation ('ml')\cite{Patwari2003},  SDP-RSS ('rss')\cite{Ouyang2010},  SDP-DISTANCE ('p-d') using pairwise distance information\cite{RN872}, SOCP-RSS ('so') modified from\cite{RN844} using RSS and SOCP-DISTANCE ('so-d') using pairwise distance information modified from \cite{Tseng2007}  to compare with our proposed estimator Robust-RSS ('ro') and several rounding algorithms ('r-r','r-g','r-p') discussed in Section III. Here 'ro' refers to the method that using the relaxation model Eq.\eqref{eq:convex form} to compute the source location without any refinement. 'r-r' refers to the combination of ro and rounding algorithm Alg.\ref{alg:rr}. 'r-g' refers to the combination of ro and rounding algorithm Alg.\ref{alg:grid searching}. 'r-p' refers to the combination of ro and rounding algorithm Alg.\ref{alg:randomization}. We assume distance information $s_i, i=1\ldots M$ of p-d and so-d is obtained from TOA measurement. Considering multi-path and NLOS transmission effects, the distance error variance introduced to $s_i$ is set to 0.15 \cite{7520939}. Other RSS based methods (ro,rss, ml,so,r-r,r-g,r-p) do not use the distance information. For the distance based methods (i.e.,p-d and so-d), the corresponding estimator is
\begin{equation}\label{eq:distance estimator}
\begin{split}
&\mathbf{\mathbf{x}}_P =\arg\min_{\mathbf{\mathbf{x}}}\sum_{i}(||\mathbf{\mathbf{x}}-\mathbf{\hat{z}}_i||-s_i)^2
\\&i=1,\ldots,M
\end{split}
\end{equation}

For SDP-DISTANCE method, the corresponding SDP estimator relaxed from \eqref{eq:distance estimator} is
\begin{equation}\label{eq:p-d estimator}
\begin{split}
&\mathbf{\mathbf{x}}_P =\arg\min_{\mathbf{\mathbf{x}},k}k
\\&s.t.
\\ &\mathrm{tr}(X)-2\mathbf{\hat{z}}_i'\mathbf{\mathbf{x}}+\mathbf{\hat{z}}_i'\mathbf{\hat{z}}_i-s_i<=k
\\ &\mathrm{tr}(X)-2\mathbf{\hat{z}}_i'\mathbf{\mathbf{x}}+\mathbf{\hat{z}}_i'\mathbf{\hat{z}}_i-s_i>=-k
\\&\begin{pmatrix}X&\mathbf{\mathbf{x}} \\ \mathbf{\mathbf{x}}^T& 1\end{pmatrix}\succeq 0
\\&i=1,\ldots,M
\end{split}
\end{equation}

SOCP-RSS method is relaxed  from\eqref{eq:add ausxiliary varialbe version}, \eqref{eq:expand quadratic form} as:

\begin{equation}\label{eq:socp-rss temp version}
\begin{split}
&\mathbf{\mathbf{x}}_P =\arg\min_{\mathbf{\mathbf{x}},k}k
\\&s.t.
\\&||\mathbf{\mathbf{x}}-\hat{\mathbf{z_i}}|| \leq \frac{1}{\zeta}(k{\beta_i}^2-\zeta^2)
\\&||\mathbf{\mathbf{x}}-\hat{\mathbf{z_i}}||^2+\zeta^2 \geq k^{-1}{\beta_i}^2,\ldots,M
\\&i=1,\ldots,M
\end{split}
\end{equation}

Noting that $||\mathbf{\mathbf{x}}-\mathbf{z_i}|| \gg \zeta$, Eq.\eqref{eq:socp-rss temp version} can be transformed into
\begin{equation}\label{eq:socp-rss with norm}
\begin{split}
&\mathbf{\mathbf{x}}_P=\arg\min_{\mathbf{\mathbf{x}},k}k
\\&s.t.
\\&||\mathbf{\mathbf{x}}-\hat{\mathbf{z_i}}|| \leq \frac{1}{\zeta}(k{\beta_i}^2-\zeta^2)
\\&\begin{pmatrix}k&\beta_i \\ \beta_i& ||\mathbf{\mathbf{x}}-\hat{\mathbf{z_i}}||\end{pmatrix}\succeq 0
\\&i=1,\ldots,M
\end{split}
\end{equation}
Then Eq.\eqref{eq:socp-rss with norm} can be transformed into

\begin{equation}\label{eq:socp-rss final}
\begin{split}
&\mathbf{\mathbf{x}}_P=\arg\min_{\mathbf{\mathbf{x}},k,t}k
\\&s.t.
\\&||\mathbf{\mathbf{x}}-\hat{\mathbf{z_i}}|| \leq \frac{1}{\zeta}(k{\beta_i}^2-\zeta^2)
\\&\begin{pmatrix}k&\beta_i \\ \beta_i&t(i)\end{pmatrix}\succeq 0
\\&||\mathbf{\mathbf{x}}-\hat{\mathbf{z_i}}|| \leq t(i)
\\&i=1,\ldots,M
\end{split}
\end{equation}
Clearly Eq.\eqref{eq:socp-rss final} is a Second-Order Cone Programming problem and can be solved by existing numerical methods efficiently. SOCP-DISTANCE  is obtained as

\begin{equation}\label{eq:socp-distance final}
\begin{split}
&\mathbf{\mathbf{x}}_P=\arg\min_{\mathbf{\mathbf{x}},k,t}k
\\&s.t.
\\&t(i)-\beta(i)^2 \leq k
\\&t(i)-\beta(i)^2 \geq -k
\\&t(i)\geq 0
\\&i=1,\ldots,M
\end{split}
\end{equation}

The modified ML-RSS estimator \eqref{eq:ML estimator of target with inaccuracy of anchor} is solved using MATLAB function 'lsqnonlin'.

We assume anchors and the source are located in a $1\times 1$ rectangle area. For simplicity, the communication range in each topology is not concerned. We apply a fixed link error model with equal RSS measurement noise variances $\sigma^2$ for all links. Here $\sigma^2$ is the variance of $n_i$ in Eq.\eqref{eq:RSS measurement model}.
We set $d_0=0.025$, $L_0=8$, and $\gamma=3$ \cite{Goldsmith2005}. Note that in this study, relative distance rather than absolute distance is considered since the SDPT3 solver needs numerically easy input. If the elements of the problem are large in magnitude, it may cause numerical inadequacy. In this situation, a scaling procedure is necessary.

For example, if the edge length of the deploy region is set to 400m, the optimal value of $X$ in Eq.\eqref{eq:convex form} may have elements in the $10^4$ or $10^5$ magnitude. Since the elements in $X$ are large, the minimum eigenvalue of $X-\mathbf{x}\mathbf{x}^T$ has a significantly different magnitude than that of $\begin{pmatrix}X&\mathbf{\mathbf{x}} \\ \mathbf{\mathbf{x}}^T& 1\end{pmatrix}$. So a very small magnitude negative min eigenvalue of $\begin{pmatrix}X&\mathbf{\mathbf{x}} \\ \mathbf{\mathbf{x}}^T& 1\end{pmatrix}$ with a very small magnitude can, and in this case does, correspond to a negative min eigenvalue of $ X-xx^T$ of a much larger magnitude. The solver only "knows" the constraint $\begin{pmatrix}X&\mathbf{\mathbf{x}} \\ \mathbf{\mathbf{x}}^T& 1\end{pmatrix} \succeq 0$, and therefore works to satisfy that within tolerance (but may not achieve that with Eq.\eqref{eq:convex form}). Then the "violation" in terms of the minimum eigenvalue of $X-\mathbf{x}\mathbf{x}^T$ can be significant, and SDPT3 solver will fail to give a feasible solution. The simulation should strive to get the non-zero elements of the optimal $X^*$ to be much closer to one in magnitude. Therefore,in numerical simulations we consider the relative distances to avoid the aforementioned situation. Using commercial solvers such as CPLEX \cite{cplex} may mitigate this issue by completing the scaling process automatically.

\begin{figure*}[htbp]%
\centering
\subfloat[][]{\includegraphics[width=0.35\linewidth]{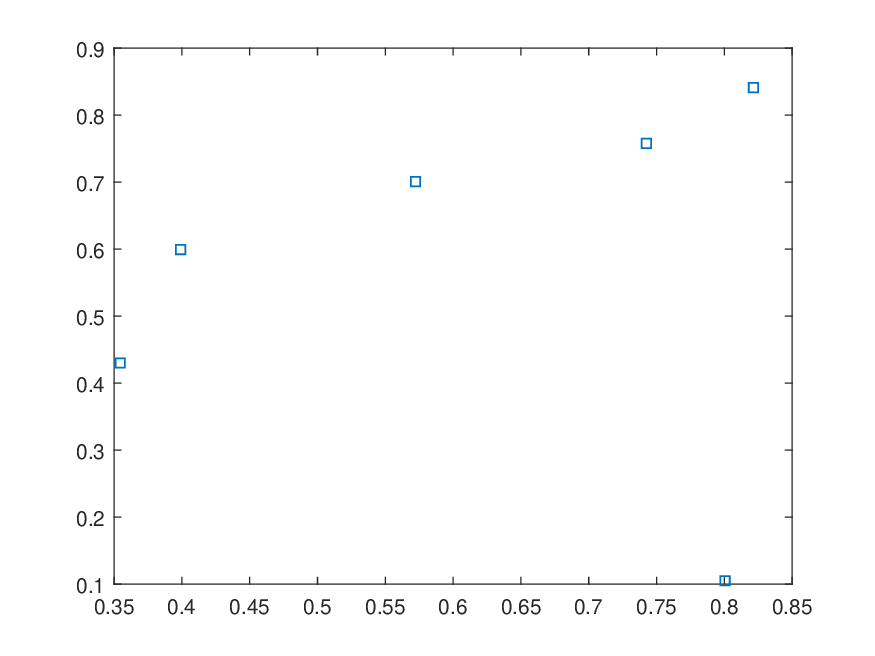}\label{sub:randomplacementanchor}}%
\qquad
\subfloat[][]{\includegraphics[width=0.35\linewidth]{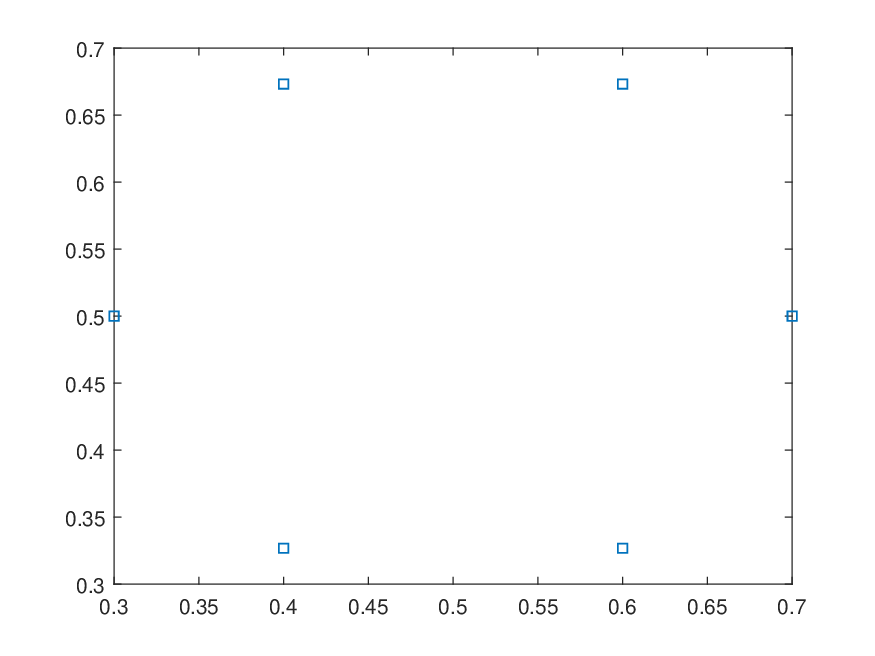}\label{sub:goodplacementanchor}}
\qquad
\caption{Two types of anchor placement. In Fig.\ref{sub:randomplacementanchor}, anchors are randomly placed. In Fig.\ref{sub:goodplacementanchor} anchors are placed by design. It is obvious that in Fig.\ref{sub:goodplacementanchor} the source will reside within the convex hull of anchors with a high probability.}%
\label{fig:two anchor placement}%
\end{figure*}

Geometric layout  has a significant impact on the localization accuracy because of the "convex hull" effect \cite{Zhang2018}.  If the distances among anchors are not sufficiently large, the localization problem can become ill-conditioned and more prone to inaccuracy. As shown in Fig.\ref{fig:two anchor placement}, two types of simulation are conducted, one with 'good' anchor placement, the other with 'bad' anchor placement (i.e., anchors deployed close to each other). Note that how to optimize geometric layout is beyond the scope of this paper, but relevant information can be found in related work \cite{wang2017optimization,wang2018improve}.

\begin{figure*}[htbp]%
\centering
\subfloat[][]{\includegraphics[width=0.35\linewidth]{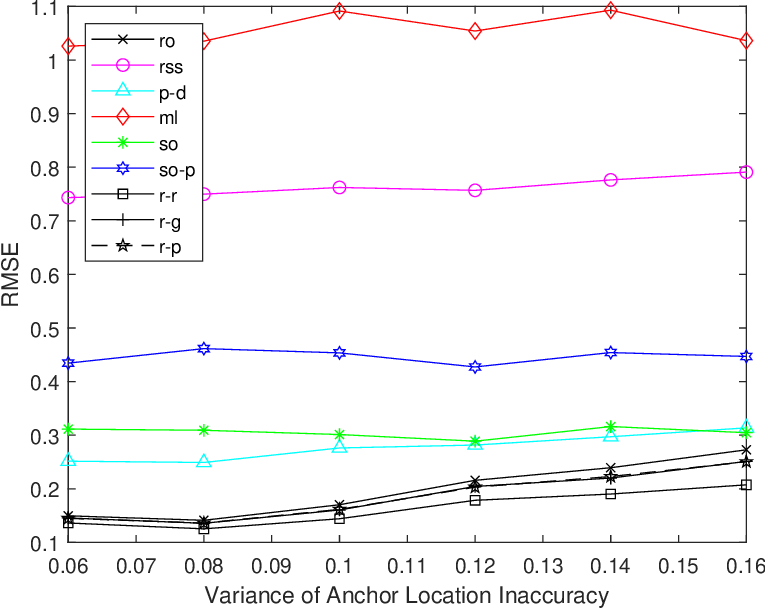}\label{sub:locerror}}%
\qquad
\subfloat[][]{\includegraphics[width=0.35\linewidth]{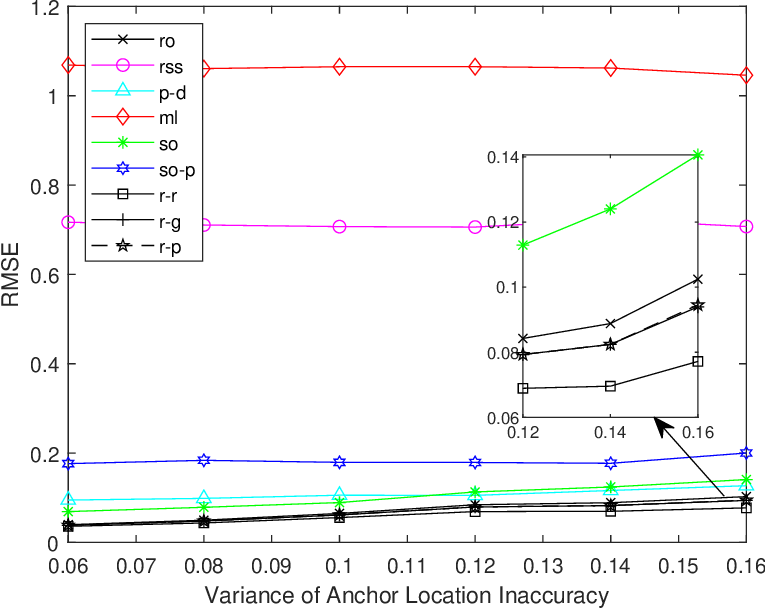}\label{sub:locerrorz1}}
\qquad
\caption{Numerical simulation results of different methods with various  value of anchor location error $\zeta$. $\sigma=0$, number of anchors $M=3$. Fig.\ref{sub:locerror} is "RMSE Versus $\zeta$" plot under 'bad' anchor placement as shown in Fig.\ref{sub:randomplacementanchor}. Fig.\ref{sub:locerrorz1} is under 'good' anchor placement as show in Fig.\ref{sub:goodplacementanchor}. 'ro' denotes the robust localization algorithm without any rounding methods. 'rss' denotes SDP-RSS\cite{Ouyang2010}. 'p-d' denotes  SDP-DISTANCE using pairwise distance information\cite{RN872}. 'ml' denotes modified ML estimation\cite{Patwari2003}. 'so' denotes SOCP-RSS modified from\cite{RN844} using RSS,'so-d' denotes SOCP-DISTANCE using pairwise distance information modified from \cite{Tseng2007}. 'r-r' is the combination of ro and rounding algorithm Alg.\ref{alg:rr}. 'r-g' is the combination of ro and rounding algorithm Alg.\ref{alg:grid searching}. 'r-p' is the combination of ro and rounding algorithm Alg.\ref{alg:randomization}. }.
\label{fig:loction error effects}%
\end{figure*}

\begin{figure*}[htbp]%
\centering
\subfloat[][]{\includegraphics[width=0.35\linewidth]{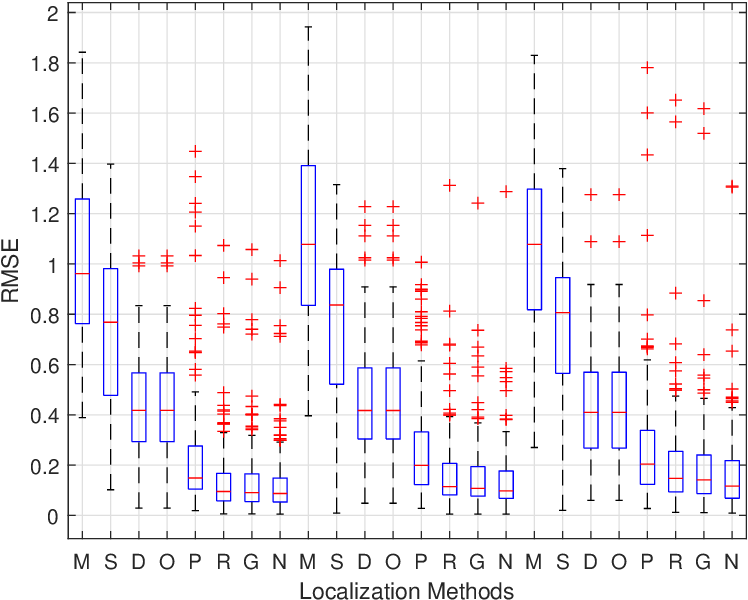}\label{sub:locerrorbox}}%
\qquad
\subfloat[][]{\includegraphics[width=0.35\linewidth]{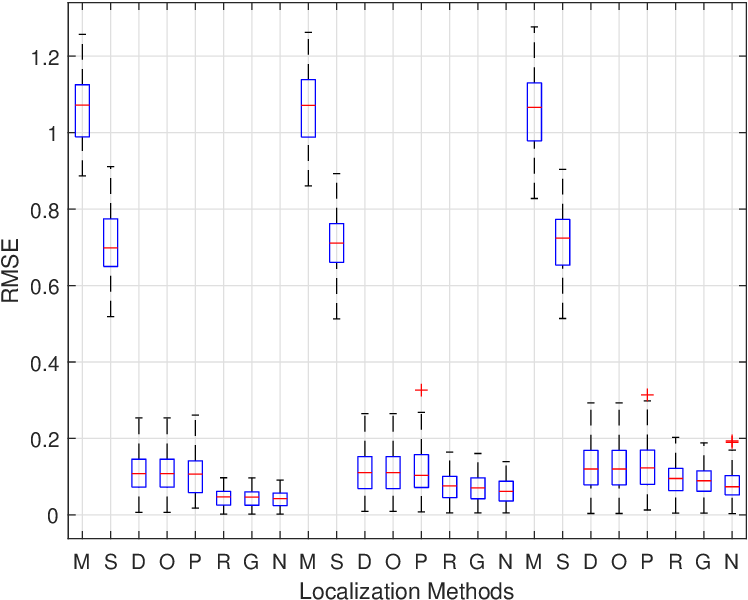}\label{sub:locerrorboxz1}}
\qquad
\caption{The boxplot of estimation errors. Fig.\ref{sub:locerrorbox} is under 'bad' anchor placement as shown in Fig.\ref{sub:randomplacementanchor}. Fig.\ref{sub:locerrorboxz1} is under 'good' anchor placement as show in Fig.\ref{sub:goodplacementanchor}. In this subplot's x-axis,'M','S','D','O','P','R','G','N' stand for 'ml','rss','p-d','ml','so-p','so','p-d','r-o','r-g','r-r' respectively. All of the following figures have the same notations. In the x-axis, the first group from 'M' to 'N' correspond to $\zeta=0.06$, the second group correspond to $\zeta=0.10$, and the third group correspond to $\zeta=0.12$.} %
\label{fig:location error boxplot}%
\end{figure*}

\begin{figure*}[htbp]%
\centering
\subfloat[][]{\includegraphics[width=0.35\linewidth]{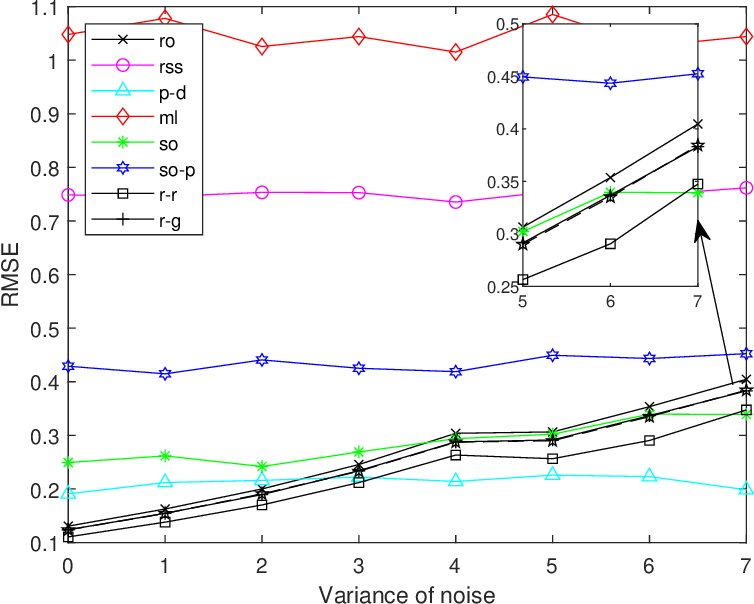}\label{sub:noise}}%
\qquad
\subfloat[][]{\includegraphics[width=0.35\linewidth]{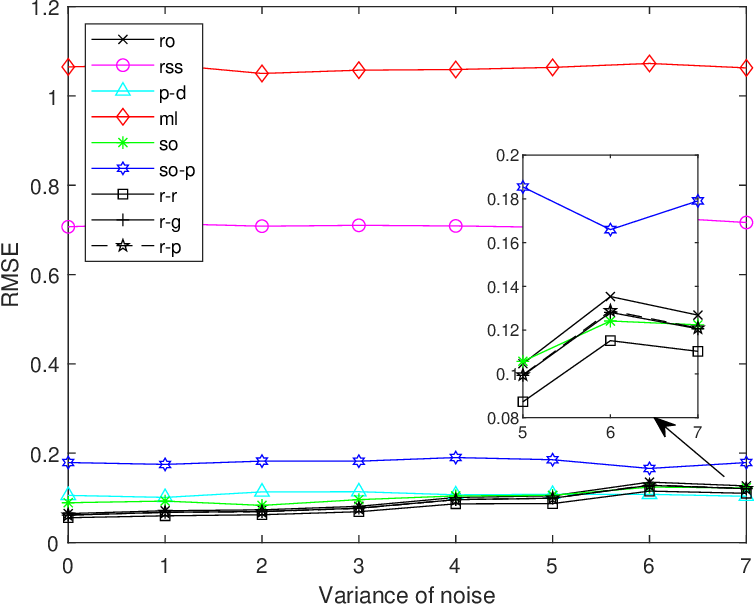}\label{sub:noisez1}}
\qquad
\caption{Numerical simulation results of different methods with various value of RSS measurement noise $\sigma$ ($\zeta=0.06$, number of anchors $M=3$). Fig.\ref{sub:locerror} is "RMSE Versus $\sigma$" plot under 'bad' anchor placement as shown in Fig.\ref{sub:randomplacementanchor}. Fig.\ref{sub:locerrorz1} is under 'good' anchor placement as show in Fig.\ref{sub:goodplacementanchor}. } %
\label{fig:nosie effects}%
\end{figure*}

\begin{figure*}[htbp]%
\centering
\subfloat[][]{\includegraphics[width=0.35\linewidth]{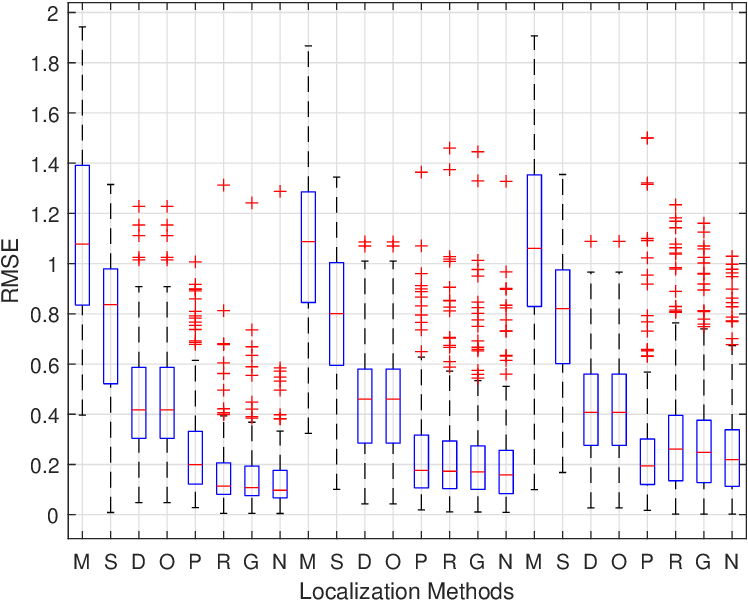}\label{sub:nbox}}%
\qquad
\subfloat[][]{\includegraphics[width=0.35\linewidth]{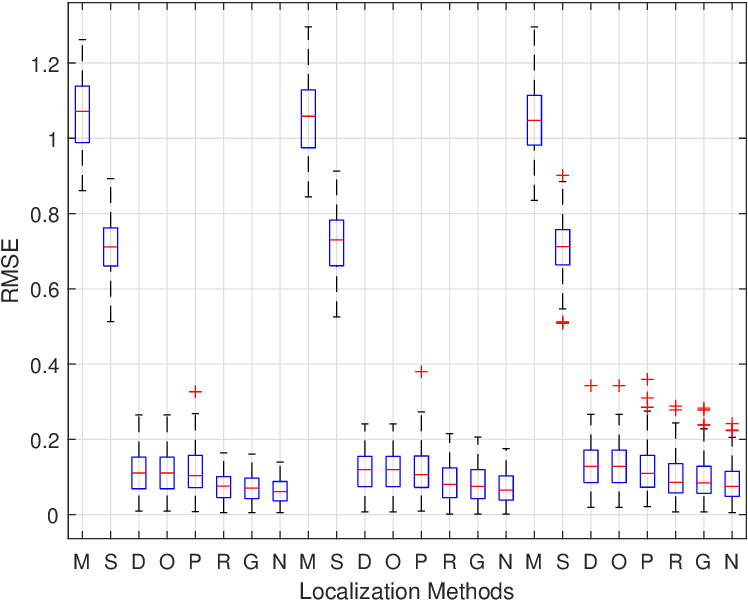}\label{sub:nboxz1}}
\qquad
\caption{The boxplot of estimation errors corresponding with different level of noise. $\zeta=0.06,M=3$. Fig.\ref{sub:locerror} is under 'bad' anchor placement as shown in Fig.\ref{sub:randomplacementanchor}. Fig.\ref{sub:locerrorz1} is under 'good' anchor placement as shown in Fig.\ref{sub:goodplacementanchor}. In the x-axis, the first group from 'R' to 'G' correspond to $\sigma=0$, the second group correspond to $\sigma=2$, and the third group correspond to $\sigma=4$.} %
\label{fig:noise effect with good anchor placement}%
\end{figure*}

\begin{figure*}[htbp]%
\centering
\subfloat[][]{\includegraphics[width=0.35\linewidth]{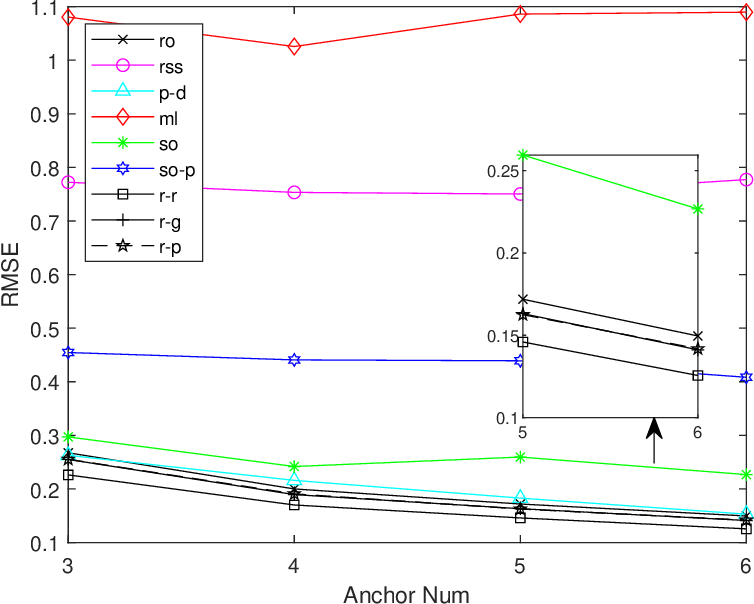}\label{sub:anchornum}}%
\qquad
\subfloat[][]{\includegraphics[width=0.35\linewidth]{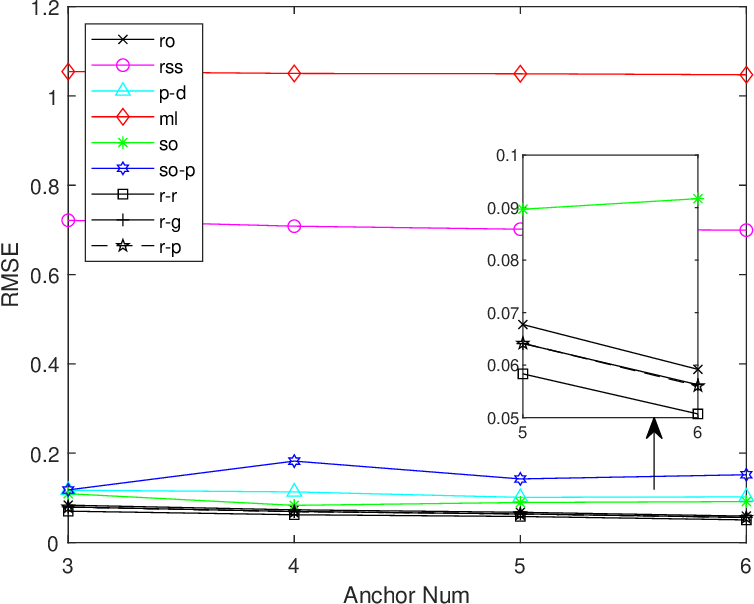}\label{sub:anchornumz1}}
\qquad
\caption{Numerical simulation results of different methods with various number of anchors $M$. ($\zeta=0.08$, $\sigma=2$). Fig.\ref{sub:anchornum} is "RMSE Versus $M$" plot under 'bad' anchor placement as shown in Fig.\ref{sub:randomplacementanchor}. Fig.\ref{sub:anchornumz1} is under 'good' anchor placement as shown in Fig.\ref{sub:goodplacementanchor}. } %
\label{fig:anchor effects}%
\end{figure*}

\begin{figure*}[htbp]%
\centering
\subfloat[][]{\includegraphics[width=0.35\linewidth]{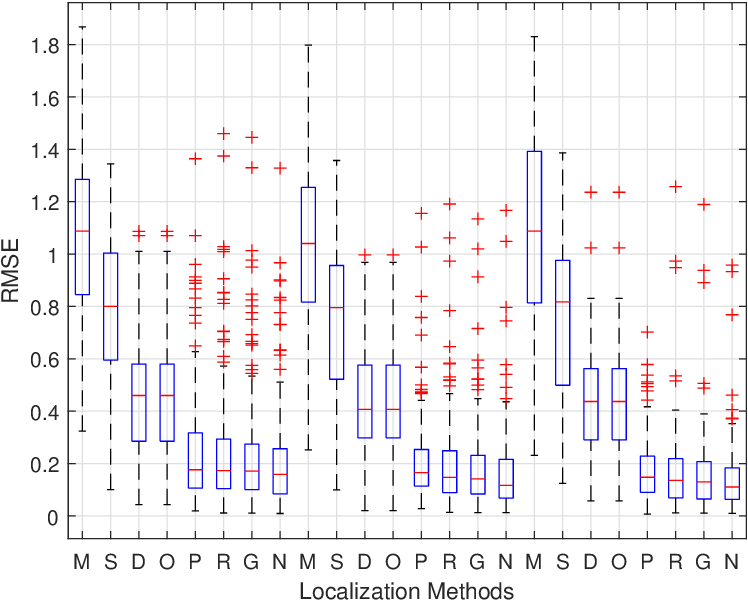}\label{sub:anchorbox}}%
\qquad
\subfloat[][]{\includegraphics[width=0.35\linewidth]{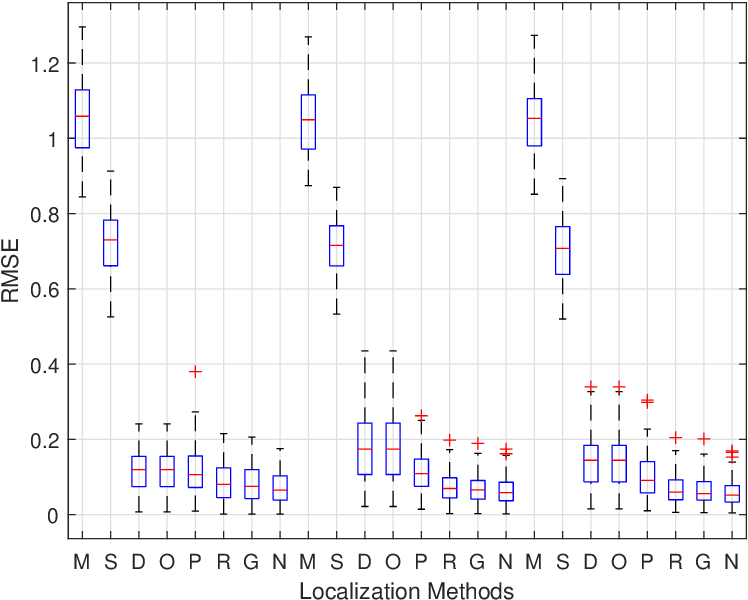}\label{sub:anchorboxz1}}
\qquad
\caption{The boxplot of estimation errors correspond to different number of anchors. Fig.\ref{sub:locerror} is under 'bad' anchor placement as shown in Fig.\ref{sub:randomplacementanchor}. Fig.\ref{sub:locerrorz1} is under 'good' anchor placement as shown in Fig.\ref{sub:goodplacementanchor}.In the x-axis, the first group from 'R' to 'G' correspond to $M=3$, the second group correspond to $M=4$, and the third group correspond to $M=5$.} %
\label{fig:anchor effects box}%
\end{figure*}

\subsection{Effects of Error Bound $\zeta$}
In this subsection, the effects of bounding error $\zeta$ is investigated, as shown in Fig.\ref{fig:loction error effects}. As the error bound $\zeta$ increases, the positions of the anchors are generally less accurate. We can observe that the performances of all methods deteriorated with an increasing error bound, but the method 'r-r' yields the best performance, due to its design for coping with model uncertainties. As shown in Fig.\ref{sub:locerror}, when the anchors are randomly deployed, 'ml' and 'rss' methods are entirely ineffective overall error level. Fig.\ref{sub:locerrorz1} shows if the deployment of anchors are carefully designed so that the source always falls into the convex hull of anchors, the so-p method gives a lower quality result. We can observe that in both Fig.\ref{sub:locerror} and Fig.\ref{sub:locerrorz1}, 'r-r' performs much better compared with 'p-d' and 'so' methods. However, in Fig.\ref{sub:locerrorz1}, 'r-r''s advantage compared with ro is not significant because the performance before rounding is sufficiently good due to the convex hull effects. It is worth noting that even in Fig.\ref{sub:locerrorz1}, where the localization accuracy is relatively high, the improvement of ro and 'r-r' methods are considerable compared with 'so' and 'p-d' estimators. When the error is relatively high ($\zeta=0.16$), there is still approximately $30\%$ of accuracy improvement.

Fig.\ref{fig:location error boxplot} is the boxplot of RMSE for different levels of anchor errors. In our case, the boxplot illustrates the distribution of the location estimation error.
If a method is robust, it demonstrates a shorter length (narrower error distribution) and lower median mark in the boxplot. For brevity, 'r-p' rounding methods is omitted since its distribution is very similar to 'r-g'.  As shown in Fig.\ref{fig:location error boxplot}, 'r-r' has superior performance in terms of lower RMSE value and narrower distribution. The distribution of 'ml' is the widest among all methods. Since the performance of 'ml' method depends heavily on the starting point, this method is impractical. There are much fewer data points outside the box in Fig.\ref{sub:locerrorboxz1} compared with Fig.\ref{sub:locerrorbox}. This phenomenon indicates another advantage of the convex hull effect.

\subsection{Effects of Measured Noise $\sigma$}
The simulation results for different values of $\sigma$ are illustrated in Fig.\ref{fig:nosie effects}.  We set $\zeta=0.06$ and $M=3$. Fig.\ref{sub:noise} is for random anchor placement, and Fig.\ref{sub:noisez1} is for 'good' anchor placement.
As can be observed in Fig.\ref{fig:nosie effects}, the performance of any RSS based methods shows degradation as $\sigma$ increases. The 'r-r' method outperforms all other methods. SDP-based methods perform better than SOCP-based methods due to the tighter relaxation. It should be noted that RSS noise will not affect distance-based methods. For 'p-d' and 'so-p', the increment of the RSS measurement noise level does not influence their performances. This is because we assume distance information are obtained from TOA rather than RSS. This is why in Fig.\ref{sub:noise} we can observe that p-d method can perform better than 'r-r'.

\subsection{Effects of the Number of Anchors $M$}
In addition to $\zeta$ and $\sigma$, the number of anchors $M$ also has an impact on the localization accuracy. The probability of the source lying in the convex hull of the anchors increases if $M$ becomes greater.
Fig.\ref{fig:anchor effects} shows the RMSE versus $M$. As expected, the RMSE is generally reduced if $M$ increases in random anchor deployments. In Fig.\ref{sub:anchornumz1}, the effects of the number of anchors is not as apparent as in Fig.\ref{sub:anchornum}. This is because the convex hull effect makes the RMSE already low.

As can be observed in both Fig.\ref{sub:anchornum} and Fig.\ref{sub:anchornumz1}, 'r-r' achieves high accuracy. For Fig.\ref{sub:anchornum}, when $M$ varies from 3 to 5, the performance improved significantly for 'ro','r-r','p-d', and 'so-p'. Fig.\ref{fig:anchor effects box} is the boxplot of RMSE of different methods corresponding with different values of $M$. As shown in Fig.\ref{fig:anchor effects box}, when $M$ increases, the width of the RMSE distribution for r-r reduces, and the median mark corresponding with r-r also declines.

\section{Conclusion and Future Works}
In this paper, the inaccurate position information of anchors in RSS-based source localization has been addressed. We propose a robust location estimator based on a min-max optimization method. This estimator considers the worst case of anchor error. Another advantage of this estimator is that we do not need to know the distribution of the error of the anchor location. We relax the estimator to an SDP problem so that it can be solved efficiently. In order to further improve the performance, we study different rounding strategies. We  propose a new rounding algorithm considering the inaccurate locations of both anchors and the source. This new rounding algorithm selects the best combination of possible anchors and the source. The selected combination achieves the 'best fit' of the RSS measurement. Simulations have evidenced that the proposed method outperforms methods in prior work that are being compared in this paper. Also, the proposed method is shown to fit well with the RSS measurement model. The influence of different environmental parameters on the localization accuracy is investigated as well. Last but not least, the performance of the proposed method is not affected by initial point for the solving process.

A further study could expand the framework established in this paper to other practical scenarios, such as wireless sensor network localization. Further research could also be undertaken to consider the self-estimation of wireless propagation parameters in the problem formulation.

\bibliographystyle{IEEEtran}

\bibliography{ieeepaper}
\end{document}